\documentclass[twocolumn,10pt,a4paper,byrevtex,floatfix,prb,unsortedaddress,superscriptaddress]{revtex4}

\usepackage[utf8]{inputenc}
\usepackage{amssymb}
\usepackage{soul}
\usepackage{graphicx}
\usepackage{hyperref}
\usepackage{subfigure}
\usepackage{amsmath}
\usepackage{color}
\usepackage[a4paper, total={7in, 9in}]{geometry}

\definecolor{DarkGreen}{rgb}{0,0.70,0}

\definecolor{Purple}{rgb}{0.5,0,0.70}

\definecolor{Blue}{rgb}{0, 0.5, 0.902}

\begin{document}

\title{Unraveling spin texture and spin-orbit coupling contributions in spin triplet superconductivity}

\author{Pablo Tuero$^{\dag}$}
\affiliation{Departamento de F\'isica de la Materia Condensada C-III, Universidad Aut\'onoma de Madrid, Madrid 28049, Spain}

\author{C\'esar Gonz\'alez-Ruano$^{\dag}$}
\affiliation{Departamento de F\'isica de la Materia Condensada C-III, Universidad Aut\'onoma de Madrid, Madrid 28049, Spain}

\author{Yuan Lu}
\affiliation{Institut Jean Lamour, Nancy Universit\`{e}, 54506 Vandoeuvre-les-Nancy Cedex, France}

\author{Coriolan Tiusan}
\affiliation{Institut Jean Lamour, Nancy Universit\`{e}, 54506 Vandoeuvre-les-Nancy Cedex, France}
\affiliation{Department of Solid State Physics and Advanced Technologies, Faculty of Physics, Babes-Bolyai University, Cluj Napoca 400114, Romania\\
\small{$^\dag$Equally contributing authors}}

\author{Farkhad G. Aliev}
\email[e-mail: ]{farkhad.aliev@uam.es
}
\affiliation{Departamento de F\'isica de la Materia Condensada C-III, Universidad Aut\'onoma de Madrid, Madrid 28049, Spain}

%\date{November 2023}

\begin{abstract}
Over the past decade, it has been proposed theoretically and confirmed experimentally that long-range spin triplet (LRT) superconductivity can be generated in superconductor/ferromagnet hybrids either by the presence of spin textures (ST-LRT) or thanks to spin-orbit coupling (SOC-LRT). Nevertheless, there has been no theoretical or experimental investigation to date suggesting that both contributions could simultaneously exist within an experimental system. To disentangle these contributions, we present a comprehensive study of superconducting quasiparticle interference effects taking place inside a ferromagnetic layer interfacing a superconductor, through the investigation of above-gap conductance anomalies (CAs) related to MacMillan-Rowell resonances. The bias dependence of the CAs has been studied under a wide range of in-plane (IP) and out-of-plane (OOP) magnetic fields in two types of epitaxial, V/MgO/Fe-based ferromagnet-superconductor junctions with interfacial spin-orbit coupling. We observe an anisotropy of the CAs amplitude under small IP and OOP magnetic fields while remaining weakly affected by high fields, and implement micromagnetic simulations to help us distinguish between the ST-LRT and SOC-LRT contributions. Our findings suggest that further exploration of Fabry-P\'erot type interference effects in electron transport could yield valuable insights into the hybridization between superconductors and ferromagnets induced by spin-orbit coupling and spin textures.
\end{abstract}

\maketitle

\section{Introduction}

For more than a decade, the conversion of ``conventional'' singlet Cooper pairs to equal-spin triplet pairs in superconductor/ferromagnet (S/F) interfaces has been linked to inhomogeneous magnetism or spin textures (STs)~\cite{Volkov2001,Keizer2006,Khaire2010,Robinson2010}. This was the foundation for the development of new energy-efficient devices and, in particular, dissipationless cryogenic memories~\cite{Birge2019}, whose operation rules are based on superconducting spintronics~\cite{Linder2015_,Eschrig2015_}. Well before these developments, Gorkov and Rashba~\cite{Gorkov2001} pointed out that singlet and triplet Cooper pair mixing could also be created by interfacial spin-orbit coupling (SOC) at the surface of a superconducting layer. More recently, a number of theoretical papers showed that, in S/F heterostructures, SOC could provide singlet to equal-spin triplet pair conversion~\cite{Bergeret2013,Bergeret2014,Jacobsen2015}, potentially producing long-range triplet (LRT) proximity effects, which was later confirmed experimentally~\cite{Niladri2018,Martinez2020,Gonzalez-Ruano2020,Gonzalez-Ruano2021,Jeon2020,Cai2021}. Importantly, some studies suggest that SOC can also be synthesized by moving electrons in inhomogeneous magnetic fields~\cite{Pekar1964} or through localized STs~\cite{Desjardins2019,Davier2023}, opening an alternative route to engineer hybrid quantum devices.

In clean S/F hybrids, Fabry-P\'erot interferences due to electron confinement can result in the so-called MacMillan-Rowell resonances (MRRs)~\cite{MRR1,MRR2,Tomasch}, originated as two Andreev-type and two normal reflections take place inside the ferromagnetic electrode interfacing the superconductor~\cite{deJong1995,Visani2012,Melnikov2012,Costa2022}. Tomasch resonances~\cite{Tomasch1965} (TRs) are similar interference phenomena occurring in the superconducting layer that have been recently put forward to be a probe to distinguish between conventional and topological superconductivity~\cite{Strkalj2024}. The above-gap conductance anomalies (CAs) originated by these two processes were suggested as an unequivocal experimental proof for equal-spin triplet formation in S/F heterostructures~\cite{Visani2012}. The full suppression of these CAs under large external magnetic fields pointed towards a direct link between LRT generation and spin textures~\cite{Visani2015}.

Recently, epitaxial V/MgO/Fe heterostructures have been proposed as an example of a S/F system coupled by interfacial SOC~\cite{Martinez2020,Gonzalez-Ruano2020,Gonzalez-Ruano2021}. These superconductor/ferromagnet hybrids showed experimental evidence for SOC-induced spin-triplet pairing, observed through the sub-gap conductance anisotropy, above-gap CAs in remanent magnetization states, and superconductivity-induced changes of the magneto-crystalline anisotropy~\cite{Martinez2020,Gonzalez-Ruano2020,Gonzalez-Ruano2021}. In this system, Rashba SOC at the Fe/MgO interface causes perpendicular magnetic anisotropy (PMA) in the interfacial atomic layers~\cite{Dieny2017}, which competes with the in-plane shape anisotropy~\cite{Martinez2018} potentially inducing interfacial spin textures. For this reason, one could expect that, in V/MgO/Fe-based junctions, both SOC and STs could contribute to spin-triplet Cooper pair formation at relatively low magnetic fields.

In this Letter, we demonstrate that studying the anisotropy of MRR-induced CAs in S/F junctions under in-plane (IP) and out-of-plane (OOP) applied magnetic fields provides a unique tool to disentangle the STs and SOC contributions to spin-triplet superconductivity. Our analysis is based on the consideration that SOC-induced LRTs (SOC-LRTs) should be more robust to magnetic field than those induced by spin textures (ST-LRTs), which should strongly diminish under a high field saturating the magnetization. We find that, in contrast to previous studies~\cite{Visani2015}, the CAs in our V/MgO/Fe-based junctions remain mostly unaffected under high applied fields, firmly pointing towards SOC-LRTs. However, at low magnetic fields (where more STs are present) the amplitude of the CAs strongly depends on the magnetic field direction. Our experimental observations are qualitatively supported by micromagnetic simulations pointing towards the anisotropic induction of spin textures in the Fe layer under low IP and OOP magnetic fields.

\section{Results and discussion}\label{section:discussion}

\begin{figure}
\begin{center}
\includegraphics[width=\linewidth]{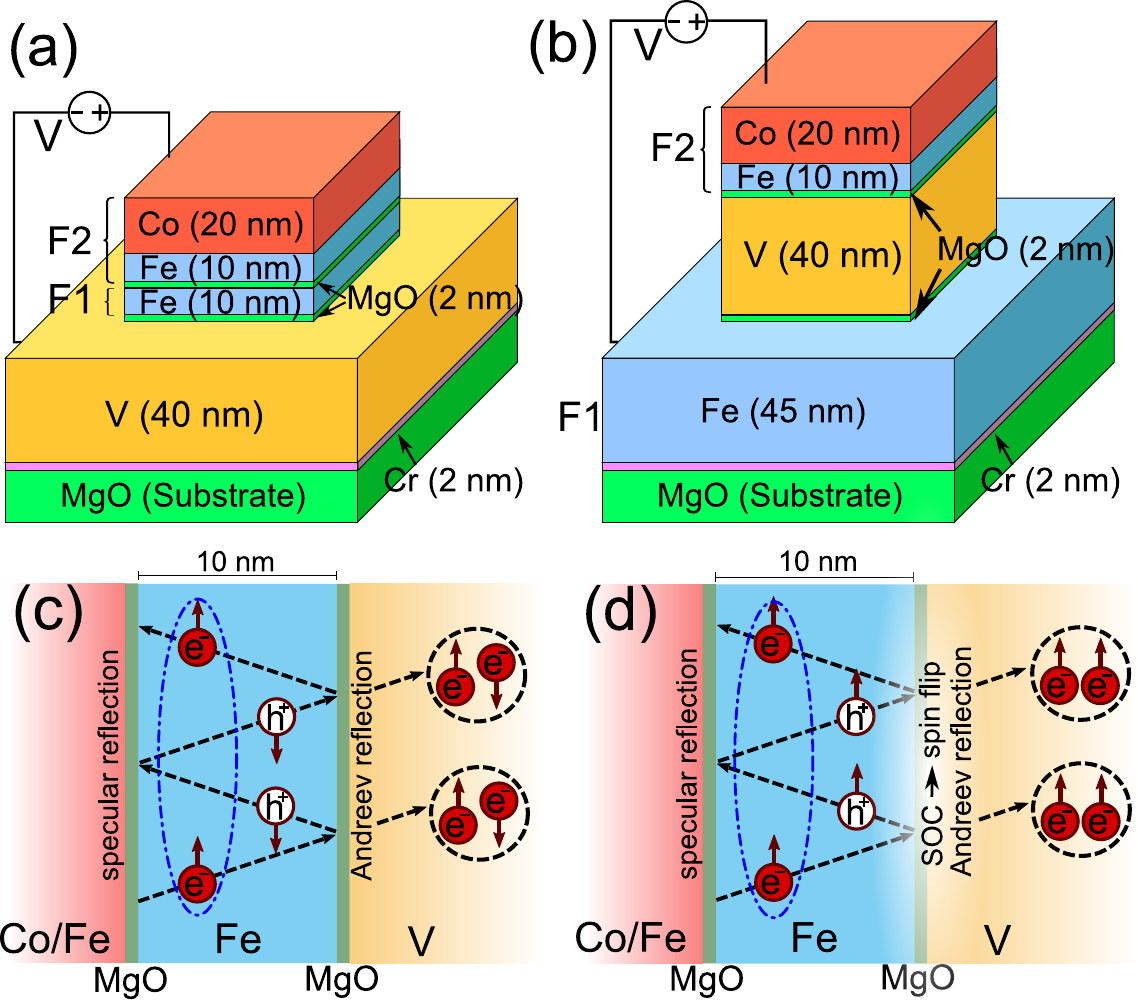}
\caption{Sketch of the S/F/F (a) and F/S/F (b) junctions, showing the disposition and thickness of each layer in the structure. The lateral sizes range from $10\times10$ to $30\times30~\mu\text{m}^2$ (samples with a larger area tend to not show such clear CAs due to interference averaging). (c)~conventional MRR, while (d) is the MRR with spin-flip, allowed in the V/MgO/Fe system thanks to interfacial SOC. The whole process is as follows: An electron in the F layer (lower part of the sketches) undergoes Andreev reflection as a hole at the interface. The reflected hole propagates to the opposite interface and is again reflected there (with a normal reflection process). It then travels back to the F/S interface and undergoes another Andreev reflection, turning again into an electron with the same spin as the initial one (upper part). Here is where the interference phenomenon takes place with the initial electron (the two electrons are circled in blue), which in turn produces the observed CAs.}
\label{SFF_sketch}
\end{center}
\end{figure}

Figure~\ref{SFF_sketch}a-b sketches the two types of samples under study.  The structure of the S/F/F junctions is V(40~nm)/MgO(2~nm)/Fe(10~nm)/MgO(2~nm) /Fe(10~nm)/Co(20~nm), while the structure of the F/S/F ones is Fe(45~nm)/MgO(2~nm)/V(40~nm)/MgO(2~nm) /Fe(10~nm)/Co(20~nm). In both cases V is the superconducting electrode, with a critical temperature ($T_C$) of about 4~K; the MgO/Fe interface provides symmetry filtering~\cite{Tiusan2007}, high spin polarization, and interfacial SOC~\cite{Lu2012}; and Fe and Fe/Co are the soft and hard ferromagnetic layers respectively. They were epitaxially grown by MBE and controlled by in-situ RHEED, as described in refs.~\cite{Tiusan2007,Lu2012} and the supplemental material. Figure~\ref{SFF_sketch}c-d illustrates the MacMillan reflections in the absence (c) or presence (d) of spin flips, induced by interfacial STs and/or SOC.

\begin{figure}
\begin{center}
\includegraphics[width=\linewidth]{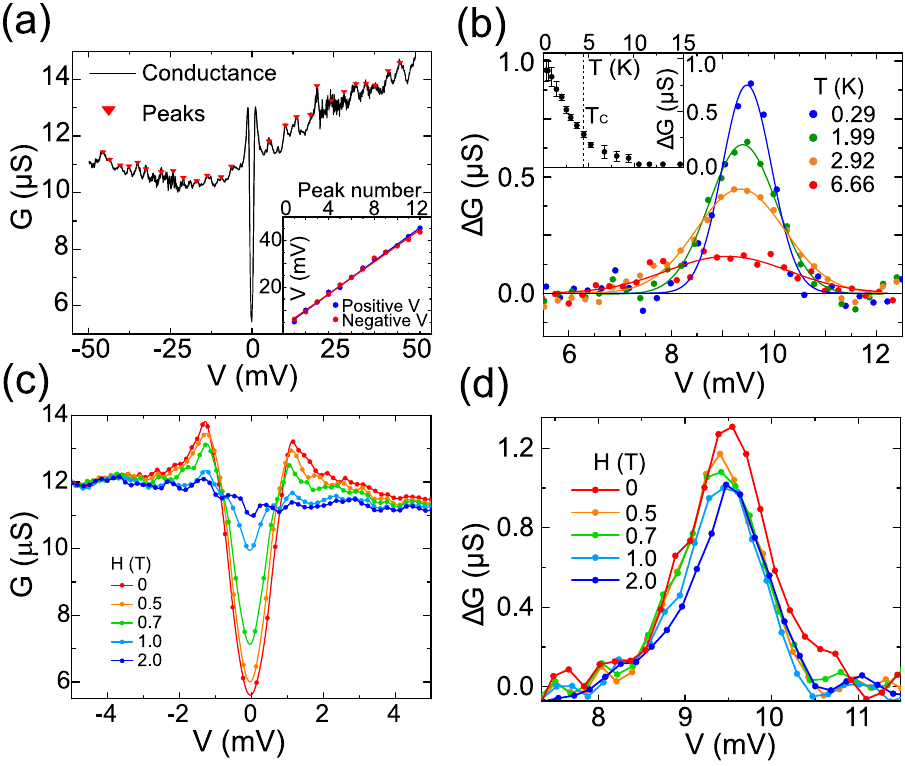}
\caption{(a)~Periodic CAs observed in a $20\times20~\mu\text{m}^2$ S/F/F sample, with red markers over each peak detected. Inset: periodicity analysis, showing peak bias vs peak number for the positive (blue) and negative (red) bias ranges with linear fits. (b)~Amplitude ($\Delta$G above the baseline conductance) of one of the CAs for different temperatures, from below to above $T_C$ ($\sim4$~K). The inset plots the amplitude of the peak above the base level vs temperature. The critical temperature is marked with a vertical dashed line. (c)~Conductance of the sample vs applied bias around the superconducting gap for different IP applied magnetic fields, showing how it disappears under an applied field of $H=2$~T. (d)~The same peak analyzed in panel (b) remains visible under high IP applied fields, even above the magnitude that suppresses the superconducting gap in the previous panel.}
\label{SFF_peak_analysis}
\end{center}
\end{figure}

Figure~\ref{SFF_peak_analysis} shows that the conductance curves of S/F/F junctions (measured at $T=0.3$~K$=0.075T_C$ unless otherwise stated) present several above-gap CAs which appear to have a periodicity with applied bias (see the inset in Figure~\ref{SFF_peak_analysis}a). As mentioned above, this kind of CAs have been linked to MacMillan and Tomasch reflections~\cite{Visani2012,Martinez2020} generating LRTs in S/F systems with highly polarized ferromagnets. However, the observed periodicity of the CAs suggests that MRRs are the leading order interference mechanism, since these reflections, depicted in Figure~\ref{SFF_sketch}c-d, give rise to conductance oscillations at periodic characteristic bias:

\begin{center}
    \begin{equation}\label{eq:peaks}
        V_n=V_0+nhv_F^\text{Fe}/4t_\text{Fe},
    \end{equation}
\end{center}

where $n=0,1,2,...$ label the successive conductance oscillation peaks, $h$ is the Planck constant, $v_F^\text{Fe}$ is the Fermi velocity in the layer where the interference process takes place (in our case, the soft Fe layer), and $t_\text{Fe}$ its thickness (10~nm in our S/F/F junctions and 45~nm in the F/S/F ones). For MRRs to occur, the gap-induced phase coherence between the initial electron and the reflected hole is maintained through the whole path back and forth the F layer (a total distance of at least 40~nm and 180~nm for the S/F/F and F/S/F junctions respectively). This is a strong hint of long-range superconducting proximity effects, as it means that superconducting correlations must survive deep into the Fe layer.

Importantly, Figure~\ref{SFF_peak_analysis}b shows how the CAs emerge mainly below $T_C$ (see the supplemental material for a detailed explanation of the CAs analysis methods). Despite this, we don't completely exclude that normal electron interference processes could also have some residual contribution. These could naturally emerge in fully epitaxial junctions due to the electron confinement (i.e. resonance transmission~\cite{Ferrari2019}) in the Fe layer, situated between two thin MgO layers acting as symmetry filters due to ``hot spots'' with enhanced electronic transmission for specific $k$ values~\cite{Butler2001}. Overall, the clear observation of CAs is indicative of the smoothness of the interfaces in the junctions ensured by the layer-by-layer epitaxial growth monitored by in-situ electron diffraction, which allows for the observation of quasiparticle interference effects. An extended discussion about the analysis of the CAs based on equation~\ref{eq:peaks}, as well as different processes that could be contributing to the conductance features in our junctions (including the above-mentioned Tomasch resonances), is given in the supplemental material.

In order to get deeper insight from the CAs, both the sub-gap conductance and one of the CAs in an S/F/F junction are thoroughly studied as a function of magnetic field in Figure~\ref{SFF_peak_analysis}c-d. Remarkably, these figures show how the superconducting gap is suppressed by an IP magnetic field of about 2~T, while the CAs remain almost unaffected by the same field.

The combination of these observations has several implications: First, the CAs are linked with the emergence of superconductivity, since they nearly disappear above $T_C$ as seen in Fig.~\ref{SFF_peak_analysis}b. Second, the fact that the CA are robust under high applied fields that suppress the superconducting gap implies that they can't be the result of singlet superconductivity, since a singlet Cooper pair with anti-parallel spins would be destroyed by such fields~\cite{Buzdin2005}. Therefore, they are most likely related to triplet superconductivity, and in particular equal-spin triplet Cooper pairs, which can survive high magnetic fields and inside a ferromagnet. Finally, the robustness of the CAs to high fields also implies that these LRTs shouldn't be exclusively produced by spin textures, as these would also be destroyed by a high applied field that fully saturates the Fe layer. Instead, we argue that the Rashba SOC at the MgO/Fe interface is the main responsible for the CAs, which nonetheless could be enhanced by an extra contribution from spin textures.

\begin{figure}
\begin{center}
\includegraphics[width=\linewidth]{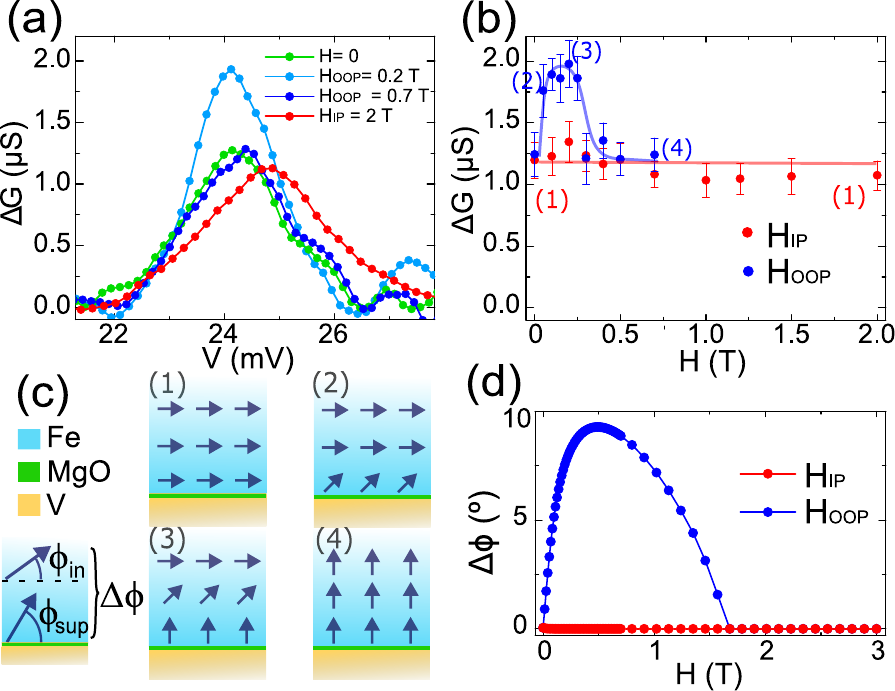}
\caption{(a)~$\Delta G(V)$ of a CA for different IP and OOP applied fields in a $30\times30~\mu\text{m}^2$ S/F/F junction after subtracting the background conductance. (b)~Amplitude of the same CA versus applied IP and OOP fields. Error bars mark the 95\% confidence bounds. The lines are guides for the eye. The numbers within the graph correspond to the different parts in panel (c), where we sketch the proposed magnetization configuration during these experiments and define the interfacial magnetization gradient (spin texture) $\Delta \phi$ simulated in the next panel. (1)~After an initial saturation with an IP field of 0.3~T, the magnetization is homogeneous. (2)~A small OOP applied field starts inducing an angle gradient in the surface layers due to the PMA. (3)~as the OOP field increases, the angle gradient reaches a maximum between the surface and inner layers. (4)~a large enough OOP field saturates the magnetization, returning the sample to a reduced interfacial spin gradient configuration. (d)~Dependence of the interfacial spin gradient $\Delta\phi$ vs IP and OOP applied field obtained from micromagnetic simulations. The supplemental material provides details about the parameters used in the simulations.}
\label{CAs_vs_H}
\end{center}
\end{figure}

To investigate the possible contribution of spin textures to the CAs and LRT formation, we have analyzed the response of the CAs to the magnetic field direction, in a field range not yet saturating the Fe layer magnetization. Figure~\ref{CAs_vs_H}a shows a $G(V)$ curve around a CA peak of another S/F/F junction at $T=0.3$~K after the background conductance is subtracted. This CA persisted for fields up to $H_\text{IP}=2$~T and $H_\text{OOP}=0.7$~T, again exceeding the critical field which suppresses the superconducting gap. The dependence of the CA amplitude with IP and OOP applied fields is presented in Fig.~\ref{CAs_vs_H}b. The CA amplitude remains approximately constant for an IP magnetic field up to 2~T, in qualitative agreement with experiments on smaller junctions. Surprisingly, when an OOP field is applied, the CA amplitude is first enhanced with respect to the IP base level. The maximum amplitude is reached for $H_\text{OOP}\approx0.3$~T, and further increasing the OOP field up to 7~kOe removes the observed CA excess. 

To understand these results, we consider the possible role of PMA-induced interfacial spin textures at the MgO/Fe interface, which is sketched in Figure~\ref{CAs_vs_H}c. It is well known that the PMA present at the MgO/Fe interfaces provides perpendicular room temperature magnetization in MgO/Fe(2~nm) films~\cite{Koo2013,Lambert2013}. In our view, the CA amplitude under IP field gives a baseline for the spin-triplet generation rate via SOC, as the magnetization of the Fe layer in this situation is nearly fully saturated thanks to the dominating IP anisotropy. By contrast, when a low $H_\text{OOP}$ is applied, the Fe atomic layers closest to the surface tend to align with this field due to the PMA, increasing the angle $\phi$ with the IP direction. The presence of these spin textures or non-collinearity close to the MgO/Fe interface could then open an additional channel for LRT generation~\cite{Bergeret2005,Robinson2010,Buzdin2007}, which in turn would enhance the CA amplitude. However, if the OOP field is further increased, the inner layers of Fe (which were initially oriented IP) would also align with the OOP magnetic field, removing the STs (therefore reducing the angle gradient $\Delta\phi$) and letting the V/MgO/Fe system again only with the interfacial SOC contribution to LRT formation.

This scenario is qualitatively supported by the results of micromagnetic simulations shown in Figure~\ref{CAs_vs_H}d, where we study the evolution of spin textures in a 10~nm thick Fe layer under IP and OOP applied fields. The simulations (see the supplemental material for details about the system and parameters used) show that, while an IP field or a large OOP field quickly saturates the magnetization, a moderate OOP field can initially maximize the angle between the magnetization of the atomic layers interfacing the MgO (where PMA is present) and the inner layers where the IP shape anisotropy dominates ($\Delta\phi$ shown in Figure~\ref{CAs_vs_H}d, the definition is sketched in Figure~\ref{CAs_vs_H}c). These interfacial STs could open the channel for ST-LRT generation, accounting for the CA enhancement under low OOP fields.

\begin{figure}
\begin{center}
\includegraphics[width=\linewidth]{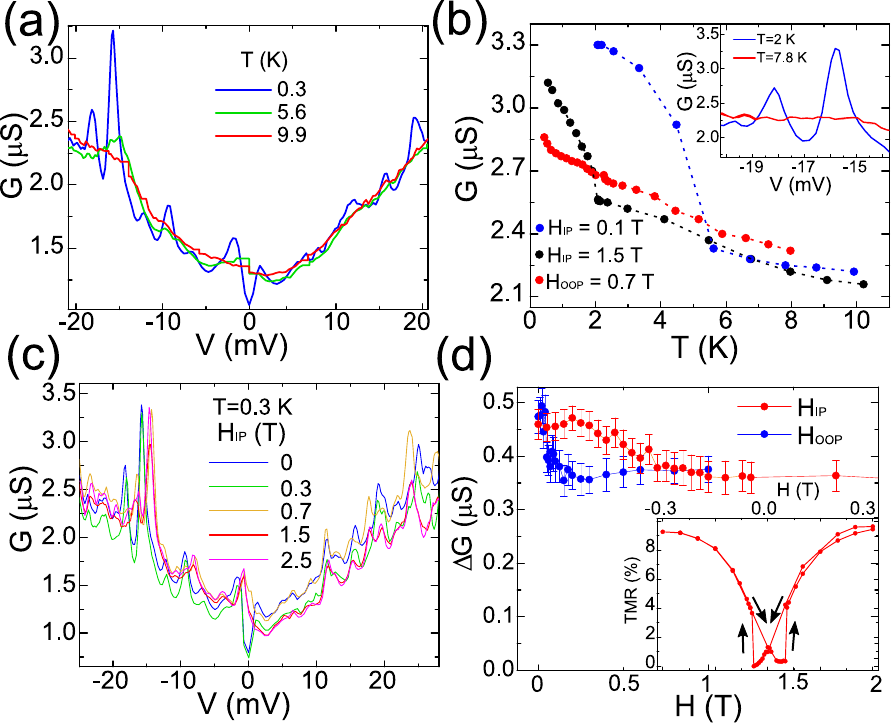}
\caption{CAs in a F/S/F junction. (a)~Conductance curves vs applied bias of a typical F/S/F junction below (blue), near (green), and above (red) the superconducting critical temperature, showing how the CAs disappear in the normal state. (b)~Maximum conductance of a selected CA near $-15$~mV vs temperature for different applied magnetic fields, both IP and OOP. The black dashed vertical line marks the critical temperature. The lines are guides for the eyes. The inset shows the conductance around the same CA below and above $T_C$. (c)~Conductance vs applied bias of the same junction, at $T=0.3$~K and for different IP applied fields, showing the robustness of the CAs under large magnetic fields. (d)~amplitude of a CA vs applied field, both IP and OOP. The inset shows an IP magnetoresistance measured at $T=0.3$~K, showing that the IP magnetization doesn't fully saturate until an applied field of about 0.3~T.}
\label{sandwich}
\end{center}
\end{figure}

We also observed similar CAs in F/S/F junctions (the structure of these samples is sketched in Figure~\ref{SFF_sketch}b), as shown in Figure~\ref{sandwich}, where the conductance oscillations found in one of these junctions are studied under different temperatures and applied magnetic fields. Here, the Fe electrode is surrounded by the substrate MgO/Cr from one side and the MgO barrier from the other one. Having in view that the Cr behaves as a symmetry-dependent metallic barrier in fully epitaxial MgO-based magnetic tunnel junctions~\cite{Greullet2007}, this Fe electrode layer will be equivalent to the 10~nm Fe layer in the S/F/F junctions. The CAs in these junctions are also robust under high applied magnetic fields, and are nearly destroyed above $T_C$ (which is slightly higher on these junctions compared to the S/F/F ones, see Figure~\ref{sandwich}b), so they are also the result of Fabry-P\'erot-like interference effects in the presence of STs and SOC. However, Figure~\ref{sandwich}d shows that the behavior of the CAs in an F/S/F junction is in some ways different to the one previously discussed in an S/F/F one. While in both types of junctions the CAs maintain a baseline level in the high field regime, in the F/S/F ones the CA enhancement at low fields is seen both for IP and OOP fields. 

We explain these differences by the influence of a much stronger anti-parallel coupling in the F/S/F junctions compared to the S/F/F ones, which could induce additional spin textures in the low IP field regime. Combined with the PMA in the atomic layers near the MgO barrier, this results in the magnetization only becoming fully saturated under an IP applied field above $0.3$~T (see Figure~\ref{sandwich}d inset, showing an IP magnetoresistance). In contrast, this extra amplitude of the CAs disappears more quickly with an OOP applied field, as in this case the PMA and the applied field point in the same direction, resulting in the magnetization being already fully saturated under a lower OOP field. As observed in the S/F/F junctions, the CAs persist in the high field regime, supporting a scenario of dominant SOC-induced LRT pair generation with an additional contribution of ST-LRTs in the low field regime.

\section{Conclusions}

In summary, our Letter demonstrates the capability to distinguish between the intrinsic (SOC) and extrinsic (spin textures) contributions to spin-triplet Cooper pair formation in superconductor-ferromagnet junctions. The method is based on the direct link~\cite{Visani2012,Martinez2020,Costa2022} already established between the emergence of periodic conductance anomalies, produced by quasiparticle interference within the ferromagnetic layer, and long-range equal spin triplet superconductivity. The anisotropic response of the CAs under different directions and magnitudes of external magnetic fields allowed us to discriminate between the two main mechanisms previously suggested to be responsible for LRT generation. Our findings call for more fundamental studies of quasiparticle interference in hybrid superconducting nanoscale systems, where superconducting correlations could be altered by the coupling of the spin and orbital motion of electrons.

\section*{Acknowledgements}
Authors thank Jacob Linder for discussions and Isidoro Martinez for help with the samples characterization. The work in Madrid was supported by Spanish Ministry of Science and Innovation (PID2021-124585NB-C32 and TED2021-130196B-C22). F.G.A. also acknowledges financial support from the Spanish Ministry of Science and Innovation through the Mar\'ia de Maeztu Programme for Units of Excellence in R\&D (CEX2018-000805-M) and ``Acci\'on financiada por la Comunidad de Madrid en el marco del convenio plurianual con la Universidad Aut\'onoma de Madrid en L\'inea 3: Excelencia para el Profesorado Universitario''.  C. T. acknowledges the UEFISCDI project ``MODESKY''ID PN-III-P4-ID- 880 PCE-2020-0230-P, grant No. UEFISCDI: PCE 245/02.11.2021.

\pagebreak

\appendix

\section*{Supplemental material}
\label{SuppMat}

\section*{Samples description and experimental set-up}

The samples were fabricated on single crystal (100) MgO substrates by molecular beam epitaxy in a chamber with $5\times10^{-11}$~mbar base pressure. The substrates were previously degassed by annealing at 600~$^{\circ}$C for 20~min before the process. Then, a 10~nm thick seed MgO underlayer was grown at 450~$^{\circ}$C on the substrate before the deposition of the other layers, to trap the segregation of any residual carbon impurities. A 2~nm Cr layer is subsequently grown to act as a seed layer for the growth of the next multilayer stack sequences. Then, each layer from the S/F/F and F/S/F junctions stacks (see Fig. 1a,b of the main text) is grown at 100~$^{\circ}$C, the metallic layers being later annealed at 450~$^{\circ}$C for 20~min to improve their surface and crystalline/epitaxial quality.

The samples were measured inside a Janis $^3$He cryostat with a base temperature of $T=0.3$~K. They have attached four contacts each: two for applying current, for which a Keithley~220 current source was used, and two for measuring voltage with a DM-552 voltimeter card integrated into the computer that controls the experiments. Several current sweeps were performed while measuring voltage in order to reduce extrinsic noise sources, and the results were averaged and differentiated to obtain the conductance.

All of the conductance vs bias measurements taken at $T=0.3$~K were performed in equilibrium conditions (i.e. with a constant base temperature). However, for the CAs amplitude vs temperature data shown in Figures~2b and 4b in the main text, the method was different. These experiments were made by initially heating the samples up to $\sim15$~K and then letting the system to slowly cool down, with a rate of about 1~K/h. During this cooling process, the G(V) curves were measured, each of them lasting around 1~h, with the temperature measured at the beginning of each curve. Due to this process, the measured temperature as shown in the figures could be overestimated by up to 0.5~K. 

\section*{Analysis methods}
\label{SuppMat:analysis}

\begin{figure}[h]
\begin{center}
\includegraphics[width=\linewidth]{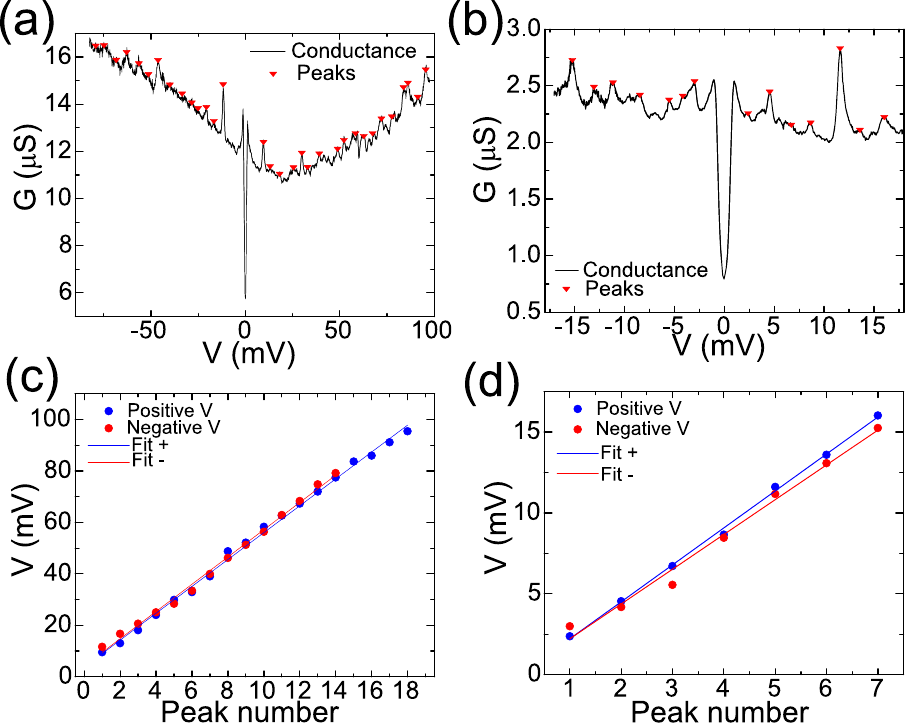}
\caption{Conductance curves for two S/F/F junctions, with $20\times20~\mu\text{m}^2$ (a) and $10\times10~\mu\text{m}^2$ (b) lateral dimensions, at $T=0.3$~K. Panels (c) and (d) show the analysis of the periodicity of the CAs for the conductance curve shown above each one respectively.}
\label{CAs_SFFs}
\end{center}
\end{figure}

\begin{figure}[h]
\begin{center}
\includegraphics[width=\linewidth]{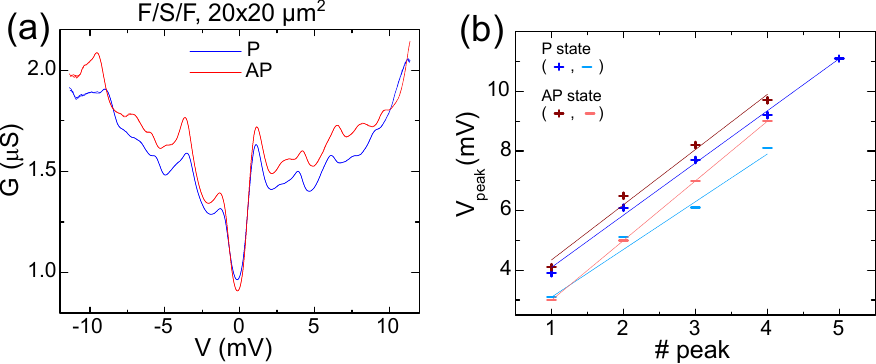}
\caption{(a)~Conductance curves for a F/S/F junction with $20\times20~\mu\text{m}^2$ lateral dimensions, at $T=0.3$~K, in the parallel (P) and antiparallel (AP) configurations of the two F layers. It's worth noting that the gap is deeper in the AP state, due to the reduced effective field from the two surrounding F layers acting on the superconductor. (b)~bias position of the above-gap conductance peaks vs peak number for positive and negative applied bias, represented with $+$ and $-$ symbols respectively, and in both magnetic configurations (dark and light blue for the parallel, dark and light red for the anti-parallel).}
\label{sandwich_P_AP}
\end{center}
\end{figure}

The conductance anomalies presented in the main text for 3 samples (two S/F/F junctions and an F/S/F one) are widely reproducible in most of the samples of the same lateral sizes that we have studied. Figures~\ref{CAs_SFFs} and \ref{sandwich_P_AP} show more examples of junctions of the two types presenting the same periodic features in the above-gap conductance.

For each sample studied, its conductance curve had to be processed to systematically find all the relevant peaks, distinguishing them from the superconducting gap and noisy features. This involved a first step where we obtained a smoothed or ``featureless'' curve, which was subtracted from the original one to remove the background trend. Once this was done, a simple peak-detecting function could be used in the desired range to find all of the relevant conductance peaks. A threshold was set for the minimum peak amplitude and voltage distance between peaks to avoid noisy features resulting in multiple detections. The results are shown in Figures~\ref{CAs_SFFs} and \ref{sandwich_P_AP}.

In order to determine the conductance height (amplitude) of each individual CA, a background conductance is subtracted by simply taking two points at each side of the studied peak (where the conductance was already flat), calculating the line that passes through them, and subtracting that baseline from the conductance in the peak range. Then, a fit to a Gaussian function was performed using the non-linear least squares method with a trust region algorithm. The peak amplitude is taken as the height of this Gaussian. Error bars mark the 95\% confidence bounds.

\section*{Extended discussion}
\label{SuppMat:V_F}

\subsection*{Fermi velocity estimation}

Based on equation~1 in the main text and the linear fits performed for the bias of the resonance peaks for different samples, an estimation of the Fermi velocity in the Fe layer can be obtained. If $m$ is the slope of the fit, then the Fermi velocity can be expressed as: 

\begin{center}
    \begin{equation}\label{eq:V_F}
        v_F^\text{Fe}=4t_\text{Fe}\cdot m/h,
    \end{equation}
\end{center}

Where to obtain the velocity in the correct units one has to multiply by the electron charge. The result is a Fermi velocity for the Fe layer of between $1.93\times10^6$ and $4.83\times10^6$~cm/s depending on the studied sample, which is somewhat low compared to tabulated values in the literature \cite{Ashcroft}. However, there are several reasons which could account for this low value.

\subsection*{Tomasch vs MacMillan resonances}

First, different processes that could also result in peaks or oscillations in the conductance, such as Tomasch reflections~\cite{Tomasch} or quantum well states, would affect our analysis process by introducing extra peak detections in the conductance curves under study. This means that there is less distance in bias between detected peaks than it should be expected with the contribution from MRRs alone, i.e. a lower slope in eq.~\ref{eq:V_F}, which results in an apparent lower $v_F$. Nonetheless, it is worth noting that the bias for the oscillations produced by Tomasch reflections is:

\begin{center}
    \begin{equation}\label{eq:TRs}
        V_n=\sqrt{\Delta^2+(nhv_F^\text{S}/2d_S)^2},
    \end{equation}
\end{center}

where $\Delta$ is the superconducting gap energy, $h$ the Planck constant, $v_F^\text{S}$ the fermi velocity in the superconducting layer, and $d_S$ its thickness. This formula is not linear in the low bias regime (for biases of about the same order of magnitude as the superconducting gap), and we don't see this deviation from a linear trend at low bias. Therefore we rule out these reflections to be the main responsible for the observed CAs in our junctions, although as stated above a small contribution could be present introducing some extra CAs.

Second, the small thickness of the layer makes the transport behavior diverge from that of the bulk. In the same line, the presence of the two MgO electrodes also affects the electronic transport, since the MgO filters out some of the $k$-vectors and different electronic symmetries present at the Fe Fermi level~\cite{Martinez2020}. In particular, the $\Delta_1$ symmetry is the most easily transmitted through the MgO barrier. Since such electrons carry less momentum on average, one could expect to have a relatively low corresponding electronic velocity. Finally, the apparently lower Fermi velocity than expected in bulk Fe could be at least in part caused by the phase delay during the Andreev reflection process as the particle and hole amplitudes penetrate into the superconductor~\cite{Eschrig2015}.

\subsection*{The role of spin-orbit coupling on transport: \textit{dips vs peaks}.}

It is also worth mentioning that, in previous observations of McMillan-Rowell resonances due to Andreev reflections in N/S~\cite{Nesher1999,Chang2004} or F/S junctions~\cite{Visani2012,Visani2015}, the conductance anomalies usually appeared as periodic peaks in the differential resistance ($dV/dI$), or conversely dips in the differential conductance ($dI/dV$). In contrast, the CAs in our V/MgO/Fe-based junctions look closer to periodic peaks in the differential conductance. We attribute this difference to the critical role that Rashba SOC plays at the MgO/Fe interface in these junctions~\cite{Martinez2020}: Due to the electronic symmetry character of the bands near the Fermi level in Fe(100), only electrons (holes) with $\Delta_1$ symmetry are transmitted from the V(100) electrode and therefore contribute to the Andreev reflection process. However, this symmetry is absent near the vanadium Fermi level. The presence of interfacial SOC allows for the conversion of $\Delta_2$ electrons into $\Delta_1$ (and vice-versa) as well as spin-flip events, making possible the Andreev reflection and therefore the transmission through the MgO barrier (along with quasiparticle penetration into the vanadium in the superconducting state) only for the electrons that have previously suffered SOC scattering. The electrons confined in the Fe electrode, which suffer multiple reflections before penetrating into the superconductor, would therefore have a strongly enhanced probability of undergoing SOC and being transmitted, resulting in a conductance peak instead of the conductance dips observed in heterostructures where SOC does not affect the conductance.
\vspace{1cm}

\begin{figure}
\begin{center}
\includegraphics[width=\linewidth]{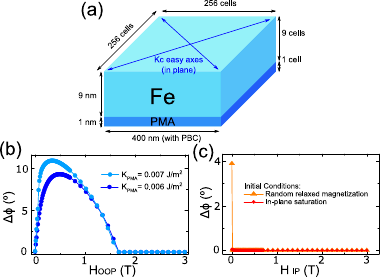}
\caption{(a)~Sketch of the system simulated in MuMax3 to study the magnetization dynamics of the Fe layer in our junctions under IP and OOP applied fields. (b)~Magnetization angle difference between the interfacial layers with PMA and the inner ones vs OOP applied field, for two values of the PMA strength parameter, showing that the behavior is qualitatively robust to variations of the chosen PMA as long as it is present. (c)~the same angle difference vs IP applied field, with an initial random relaxed magnetization and with an initial IP saturation.}
\label{mumax_simulation}
\end{center}
\end{figure}

\section*{Micromagnetic simulations}
\label{SuppMat:simulations}

Micromagnetic simulations were performed using the MuMax3 software \cite{mumax} in a thin ferromagnetic layer with PMA under the effect of an OOP magnetic field, as sketched in Figure~\ref{mumax_simulation}a. As seen in Figure~\ref{mumax_simulation}b and c, the behavior of the spin textures (measured as the magnetization angle difference between the interfacial and inner layers) is qualitatively similar to the experimental results for the S/F/F junctions: a small applied IP field quickly saturates the magnetization when the initial state is a random magnetization. On the other hand, interfacial spin textures are absent when the magnetization is initially saturated by a 0.1~T field (the experiments were done with an initial IP saturation). However, an OOP field first induces a maximum angle difference before saturating the magnetization under a sufficiently high OOP field.
The parameters used for the simulated Fe were a damping parameter $\alpha=0.02$; saturation magnetization $M_{\text{sat}}=1.7~\text{T}$; an exchange stiffness constant $A_{\text{ex}}=21\times10^{-12}~\text{J/m}$; a cubic anisotropy $K_{\text{C}}=4.8\times10^{4}~\text{J/m}^{3}$; with a number of cells equal to $N_X=N_Y=256$; $N_Z=10$ corresponding to dimensions $L_X=L_Y=400~\text{nm}$; $L_Z=10~\text{nm}$. The PMA was set as an uniaxial anisotropy in the OOP direction with anisotropy constant $K_U=K_{PMA}\times~N_Z/L_Z$, which is shown in Figure~\ref{mumax_simulation}b for two different strengths to prove that the observed behavior is qualitatively robust and not a result of fine-tuning.

\end{document}